# DNA transport is topologically sculpted by active microtubule dynamics

Dylan P. McCuskey[a], Raisa E. Achiriloaie[a], Claire Benjamin[a], Jemma Kushen[a], Isaac Blacklow[a], Omar Mnfy[a], Jennifer L. Ross[b], Rae M. Robertson-Anderson[c] and Janet Y. Sheung[a]*

[a]Department of Natural Sciences of Scripps and Pitzer Colleges, 925 N Mills Ave, Claremont, CA, 91711, USA.
[b]Syracuse University Department of Physics, Crouse Dr, Syracuse, NY, 13210, USA.
[c]University of San Diego Department of Physics and Biophysics, 5998 Alcala Park, San Diego, CA, 92110, USA.
**Email:** *jsheung@natsci.claremont.edu.

**This PDF file includes:**

> Main Text
> Figures 1 to 5

**Abstract**

The transport of macromolecules, such as DNA, through the cytoskeleton is critical to wide-ranging cellular processes from cytoplasmic streaming to transcription. The rigidity and steric hindrances imparted by the network of filaments comprising the cytoskeleton often leads to anomalous subdiffusion, while active processes such as motor-driven restructuring can induce athermal superdiffusion. Understanding the interplay between these seemingly antagonistic contributions to intracellular dynamics remains a grand challenge. Here, we use single-molecule tracking to show that the transport of large linear and circular DNA through motor-driven microtubule networks can be non-gaussian and multi-modal, with the degree and spatiotemporal scales over which these features manifest depending non-trivially on the state of activity and DNA topology. For example, active network restructuring increases caging and non-Gaussian transport modes of linear DNA, while dampening these mechanisms for rings. We further discover that circular DNA molecules exhibit either markedly enhanced subdiffusion or superdiffusion compared to their linear counterparts, in the absence or presence of kinesin activity, indicative of microtubules threading circular DNA. This strong coupling leads to both stalling and directed transport, providing a direct route towards parsing distinct contributions to transport and determining the impact of coupling on the transport signatures. More generally, leveraging macromolecular topology as a route to programming molecular interactions and transport dynamics is an elegant yet largely overlooked mechanism that cells may exploit for intracellular trafficking, streaming, and compartmentalization. This mechanism could be harnessed for the design of self-regulating, sensing, and reconfigurable biomimetic matter.

**Significance Statement**

Macromolecular transport through the cytoskeleton is critical to wide-ranging cellular processes from streaming to replication. Steric hindrances and rigidity imparted by this filamentous network often trigger anomalous subdiffusion, while active processes may induce superdiffusion. The interplay between these antagonistic contributions remains elusive. We discover non-gaussian, multi-modal transport of linear and circular DNA through kinesin-driven microtubule networks that depends non-trivially on motor activity and DNA topology. Circular DNA exhibits both enhanced superdiffusion and subdiffusion compared to linear DNA, indicative of microtubule threading that confers both directed transport and stalling. Cells may leverage these topological interactions to sculpt transport in myriad processes.

**Introduction**

The transport of macromolecules through biological cells is critical to wide-ranging cellular processes including cell signaling, development, and transcription [1–4]. It is a grand challenge of biophysics, cell biology, and bioengineering to understand transport phenomena in complex, aqueous environments. The cytoskeleton, a network of interpenetrating filamentous proteins, such as rigid microtubules, provides structural support to cells and can cage and hinder the thermal diffusion of macromolecules [5–9]. At the same time, active processes, such as motor-driven contraction and sliding of filaments, allows the cytoskeleton to dynamically restructure, move and generate forces [7,10,11]. These energy-dissipating dynamics, in turn, can cause surrounding macromolecules to undergo athermal motion via hydrodynamic, steric or chemical coupling to the network. For example, kinesin motors walking along microtubules can provide directed transport of cargo and contribute to cytoplasmic streaming and advective mixing, processes that allow macromolecules to evade the constraint of slow thermal diffusion [4,12].

Athermal evasion tactics are especially important for large molecules that are hindered by the steric constraints of the cytoskeleton. For example, active transport of DNA or RNA is required for viral infection, chromosomal compaction, Hedgehog signaling, and effective gene therapy methods [13–17]. Here, we discover that the topology of the DNA polymer has a dramatic effect on its ability to leverage the active dynamics of the cytoskeleton to accelerate transport.

Anomalous subdiffusion of large molecules is expected due to the steric hindrances and viscoelasticity of the cytoskeleton. Anomalous diffusion is characterized by motion of a particle that does not adhere to normal Brownian diffusion in which the mean-squared displacement (MSD, $\langle(\Delta r)^2\rangle$) scales linearly with lag time $\Delta t$. where $\alpha$ is the anomalous scaling exponent and $\alpha < 1$. This subdiffusion can arise from a number of different mechanisms including caging, coupling to the viscoelastic relaxation of the environment, pinning, and heterogeneous step sizes [18,6,19,20]. For example large DNA molecules have been shown to become entangled with and caged by reconstituted cytoskeleton networks; circular DNA may even become threaded and pinned by filaments, nearly halting their motion [6,19]. The mechanisms underlying the myriad observations of subdiffusion within in vivo and in vitro crowded cell-like environments, and the spatiotemporal scales over which distinct mechanisms contribute to the dynamics, remains a topic of fervent investigation [21–33].

This topology-dependent coupling bolsters numerous previous studies that have demonstrated the strong effect that DNA topology has on its diffusive dynamics in entangled and crowded environments [19,34–38]. For example, the primary mode of diffusion for linear DNA within an entangled polymer network is reptation, whereby the center-of-mass motion of the DNA chain is confined to a direction parallel to its contour, such that it diffuses 'head-first' through the surrounding network entanglements [39,40]. Circular DNA lacks the free ends necessary for this 'head-first' diffusion, so must adopt alternative diffusive mechanisms, such as modified reptation in which rings adopt amoeba-like conformations with branches or loops that can reptate [36,41]. The endless conformation of rings can also cause rings within linear polymer networks to become threaded by surrounding linear chains which essentially pins them in place until the threading chain diffuses out of the center of the ring [42–44], thereby releasing its constraint. This constraint release process is much slower than reptation or modified reptation, which can lead rings within crowded and entangled polymer networks to exhibit much slower diffusion [45], more pronounced subdiffusion [41,45], and even bi-phasic transport [19] that arises from the limitation of a threaded ring to diffuse beyond a distance comparable to its radius of gyration $R_G$. These features have recently been reported for ring DNA diffusing in in vitro cytoskeleton networks and shown to be absent for linear DNA diffusing in the same networks [19]. Linear DNA transport has instead been shown to be governed by reptation as well as caging within and hopping between the pores of the network.

How these distinct mechanisms and transport features are impacted by active restructuring of the surrounding network remains an important open question, critical to understanding the diversity of intracellular transport behaviors and processes that they govern. Moreover, how the different modes of coupling of macromolecules to the pervading network (e.g., entanglements, threadings, caging) can be leveraged to evade constraints and harness the active dynamics of their surroundings can provide powerful new mechanistic insight into complex cellular processes such as corralling, mixing, localization, condensation, phase separation, and flow. Here, we address this grand challenge by investigating the transport of large linear and ring DNA molecules within kinesin-driven microtubules networks with and without the chemical fuel (ATP) that allows kinesin motors to walk along and pull on neighboring microtubules. We discover that ring DNA couples much more strongly to the microtubule network than linear DNA, providing a direct route towards elucidating distinct contributions to transport and determining the impact of coupling on the transport signatures.

We use single-molecule tracking to show that transport of both circular and linear DNA molecules, within networks of microtubules crosslinked by kinesin motors, is non-Gaussian and multi-modal. The degree and spatiotemporal scales over which these features manifest depend on the DNA topology and state of kinesin activity. We discover that ring DNA exhibits either markedly more pronounced subdiffusion *or* enhanced superdiffusion, as compared to their linear counterparts, in passive versus active networks. Moreover, the impact of activity on linear chains is much weaker and surprisingly in opposition to rings, leading to subdiffusive, non-Gaussian transport features for linear chains that are absent in passive networks. Finally, the critical lengthscales that govern the different transport modes are indicative of microtubules threading ring DNA, which leads to both stalling and advection, while linear chains undergo a combination of caging, hopping and free diffusion tuned by the kinesin activity.

It is rather remarkable that these substantial differences in transport of ring and linear DNA arise from a single connection of the two free ends. Both topologies are 115 kilobasepairs (kbp) DNA chains comprising ~770 persistence lengths and adopting random coil configurations with radii of gyration that differ by less than a factor of two ($R_{G,L} \simeq 960$ nm and $R_{G,R} \simeq 580$ nm) [46]. We show that the simple act of cutting a ring at a single location can essentially wipe out its ability to couple to the surrounding network. While evidence of threading of flexible ring polymers by their flexible linear counterparts has been discussed extensively in the literature [36,42–44,47–49] and is now a generally accepted mode of ring transport, the ability of rings to be threaded by relatively thick and rigid filaments such as microtubules, ~25 nm in diameter, is a nontrivial result that has yet to be appreciated. More generally, leveraging macromolecular topology as a route to programming macromolecular interactions and transport dynamics represents an elegant yet largely overlooked phenomenon that cells may exploit, and that can be harnessed for the design of self-regulating, sensing, and reconfigurable active matter.

## Results and Discussion

**Kinesin-driven microtubule networks and DNA topology dictate transport properties of linear or ring DNA.** We investigate the transport of large linear and circular DNA molecules through in vitro microtubule networks that are either passively crosslinked or actively strained by kinesin motors (Fig 1A-C). We use clusters of kinesin motors that can link together two neighboring microtubules, and, when fuelled by ATP, walk along each filament such that two anti-aligned filaments can contract towards or slide past one another. These clusters have been used extensively to create flows in active nematics formed by depleted bundles of microtubules [50–63]. In those systems, the bundles are not entangled with one another and the fluid-like gliding motion of the large bundles dominates the system dynamics. Here, the network comprises single entangled microtubules that are randomly oriented and have a nominal mesh size of $\xi \simeq 1.1$ μm. The motor concentration is set to have a motor:tubulin ratio of $R \simeq 0.037$ which results in a lower bound length between motor-linkages of $l \simeq 34$ nm (see Methods), assuming all motors are bound at any point in time. This small length, relative to the mesh size, ensures that the network dynamics are dominated by either passive crosslinking or active straining, with contributions from steric entanglements, filament bending, and quiescent fluctuations playing minimal roles. This simplification allows us to clearly delineate the passive and active contributions of the network to the DNA transport.

The stark differences previously observed between the transport of ring and linear DNA in steady-state cytoskeleton composites, with evidence of threading of rings that nearly halted motion [6], motivated our examination of the topology dependence of DNA transport in active systems. Namely, we sought to determine if activity enhances or suppresses the differences between ring and linear DNA transport, and if the distinct topology-dependent coupling of the DNA to the network could withstand active restructuring. The previous evidence of threading was shown for composites of actin and microtubules, and it was assumed that most of the threading events were from actin filaments, which are ~4x thinner than microtubules (~6 nm vs ~25 nm) so could presumably more easily penetrate the random coil rings that have a radius of gyration of $R_{G,C} \simeq 580$ nm [46,64]. Moreover, in these previous experiments, the filaments were polymerized in the presence of the DNA so they could likely become more easily threaded and pinned as the filaments polymerize and the network forms. Here, our network comprises only microtubules and kinesin and the networks are formed prior to adding DNA to the system. These experimental design features allow us to answer the questions: How robust is the strong topology dependence of DNA transport in cytoskeletal networks? How does network activity alter the fingerprint of DNA transport and its dependence on topology? Can DNA topology be harnessed as a mechanism for mediating transport efficiency through the cytoskeleton, and as a facile read-out of network dynamics?

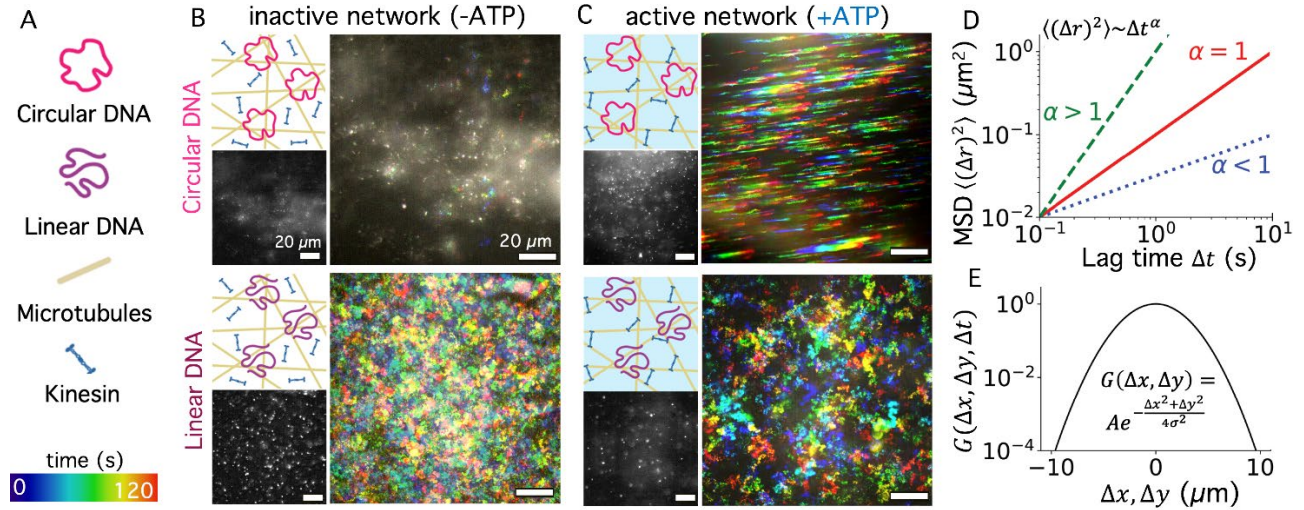

**Figure 1. Kinesin-driven microtubule networks dictate transport of linear and circular DNA.** (A) We examine the transport of fluorescent-labeled 115-kbp DNA of circular (pink) and linear (purple) topologies embedded in networks of entangled microtubules (yellow) that are passively crosslinked (B, inactive) or actively driven (C, active) by dimers of kinesin motors (dark blue) in the absence (white background, B) or presence (blue background, C) of ATP. (B,C) We record epifluorescence microscopy videos of fluorescent-labeled circular (top) and linear (bottom) DNA in inactive (B) and active (C) samples. Shown are cartoons (top left), single images (bottom left) and temporal color maps (right) depicting the colorized motion of the DNA over the course of a sample video, with color indicating time according to the scale. (D) Idealized MSDs versus lag time $\Delta t$, described by $\langle(\Delta r)^2\rangle \sim \Delta t^\alpha$, for particles undergoing normal Brownian diffusion ($\alpha = 1$, red solid), superdiffusion ($\alpha > 1$, green dashed), or subdiffusion ($\alpha < 1$, blue dotted). (E) Idealized van Hove distribution $G(\Delta x, \Delta y, \Delta t)$ for a given lag time

$\Delta t$ described by a Gaussian function (equation shown), as observed for molecules undergoing normal Brownian motion.

To characterize the transport of the DNA through the networks, we track single fluorescent-labeled DNA molecules embedded in the networks (Fig 1B-E). The temporal color maps shown in Fig 1B,C indicate the relative degree to which the molecules move over the course of a single video for the four different cases we examine: ring versus linear DNA embedded in passive versus active networks (Fig 1B,C). Both the DNA topology and the activity state clearly have an impact on the mobility. Rings appear to switch between halted motion to directed transport in passive versus active networks, while linear DNA appears to have a weaker dependence on activity and actually become less mobile upon activity. To quantify the transport visualized in Fig 1B,C, we examine the mean-squared displacements (MSDs) of the DNA (Fig 1D) and the distribution of displacements, i.e., van Hove distributions (Fig 1E).

As discussed in the Introduction, particles undergoing normal Brownian motion exhibit mean-squared displacements that scale linearly with lag time and are equal in all directions, $\frac{1}{3}\langle(\Delta r)^2\rangle = \langle(\Delta x)^2\rangle = \langle(\Delta y)^2\rangle = \langle(\Delta z)^2\rangle = 2D\Delta t$, where the constant of proportionality is the diffusion coefficient $D$ (Fig 1D). For many biological or biomimetic systems, this Brownian assumption breaks down and the particles instead exhibit anomalous diffusion described by a nonlinear scaling relation $\langle(\Delta r)^2\rangle = 2K\Delta t^\alpha$ where $K$ is a generalized transport coefficient and $\alpha < 1$ or $\alpha > 1$ signify subdiffusion or superdiffusion, respectively (Fig 1D). The former is often observed in crowded, viscoelastic and/or confined environments [23,65], and the latter can arise from active transport or advection [1,5]. Another important indicator of transport properties is the van Hove distribution of displacements evaluated for different lag times. Normal Brownian motion results in a Gaussian distribution centered at zero and with a variance $\sigma^2$ that relates to the diffusion coefficient (Fig 1E). However, for many biological and biomimetic systems, the distributions exhibit non-Gaussian features such as exponential large-displacement tails that extend beyond the Gaussian profile, as well as larger than expected modes due to more near-zero displacements [8,66,9]. The former has been attributed to hopping or jumping between pores of a network, while the latter is often a result of the surrounding network 'caging' or 'trapping' molecules [5,19,9]. Finally, for particles undergoing directed motion, the distribution may peak at a non-zero value $\Delta x_p$, indicating a most probable speed, $v_p \approx \Delta x_p/\Delta t$, with $\Delta x_p$ increasing with increasing lag time. We will consider all of these features and their plausible mechanisms when evaluating our data presented below.

**Circular DNA exhibits multimodal subdiffusion in quiescent microtubule networks that is distinct from normal diffusion of linear chains.** We first examine the transport characteristics of linear and ring DNA in quiescent microtubule networks that lack ATP such that the kinesin clusters act as passive crosslinkers (Fig 1B). We observe starkly different scaling and magnitudes of MSDs for linear and ring DNA (Fig 2A). Linear DNA exhibits nearly normal diffusion for all measured lag times, with $\alpha \approx 1$ and an effective diffusion coefficient of $D \simeq 1.19$ µm² s⁻¹. Conversely, $\langle(\Delta r)^2\rangle$ for ring DNA is lower in magnitude and much more subdiffusive than for linear DNA, with $\alpha \approx 0.5$ for $\Delta t \lesssim 1$ s followed by a crossover to a region with an unphysical negative dependence on lag time for $\Delta t \gtrsim 1.2$ s (Fig 1A). We hypothesize that this downturn may arise from multiple populations of transport modes that persist for different lengths of time. Specifically, if there were a slow mode in which molecules were nearly halted then many of them would remain in the field of view for much longer than for fast moving particles, so each of these molecular trajectories or 'tracks' would contribute equally to $\langle(\Delta r)^2\rangle$ at all evaluated lag times ($\Delta t \leq \Delta t_{max} = 2$ s). Faster moving molecules may diffuse laterally out of the 50 µm x 50 µm field-of-view (FOV) or vertically out of the ~1 µm focal depth at shorter times, so they may have comparatively shorter tracks ($T < \Delta t_{max}$) that would be unable to contribute to the ensemble-averaged MSD

at longer lag times that exceed their track duration $T$. A decrease in the contribution from fast modes with increasing lag times would result in an apparent downturn in $\langle (\Delta r)^2 \rangle$.

To test this hypothesis, we examine the distributions of track durations $T$ for circular and linear DNA (Fig 2B), which clearly show an extended tail of long-$T$ tracks for the circular DNA that is absent for linear DNA. To determine the contributions from the shorter tracks versus the extended tracks we define the track duration at which the distribution for circular DNA reaches $e^{-1}$ of its mode (peak) value as a critical time $T_c$ that divides the data into 'short' and 'long' tracks. Fig 2C,D show separate MSDs computed from tracks with $T \leq T_c$ versus $T > T_c$, which we scale by the lag time such that a horizontal line is indicative of normal diffusion, i.e., $\langle (\Delta r)^2 \rangle / \Delta t = 2D\Delta t^0$ where the y-intercept is $2D$. Subdiffusion results in negative power-law scaling, $\langle (\Delta r)^2 \rangle / \Delta t = 2K\Delta t^{\alpha-1}$. Linear DNA exhibits statistically indistinguishable MSDs for both classes of tracks, which both exhibit largely normal diffusion with $D \simeq 1.15$ µm² s⁻¹ (Fig 2C), aligning with the value we measured from the MSD from the full distribution of tracks (Fig 2A). These results suggest that there are minimal steric interactions between the microtubule network and the linear DNA that slow or otherwise alter the free diffusion of the DNA. This effect may be due to the radius of gyration of the linear DNA being slightly smaller than the network mesh size ($R_{G,L} \simeq 960$ nm versus $\xi \simeq 1.1$ µm), so confinement effects by the surrounding network are limited [8,67–69].

In stark contrast, MSDs for circular DNA exhibit varying degrees of subdiffusion, with long and short tracks producing MSDs with very different magnitudes and scalings. Short tracks display subdiffusive transport with scaling exponent of $\alpha_s \simeq 0.7$ for all lag times, while long tracks display two phases of steeper power-scaling with exponents of $\alpha_{l,1} \simeq 0.50$ for $\Delta t_c \lesssim 0.6$ s and $\alpha_{l,2} \simeq 0.20$ for $\Delta t_c \gtrsim 0.6$ s. This crossover time $\Delta t_c$ is similar to the time at which the aggregate MSD begins to roll over (Fig 1A). These results indicate that there are indeed at least two modes of anomalous transport that contribute uniquely to the dynamics of circular DNA with the faster of the two not contributing at longer lag times.

To further corroborate this finding, we plot the number of short ($T \leq T_c$) and long ($T > T_c$) tracks $N_s$ and $N_l$ that contribute to data for each lag time, relative to the total tracks at $\Delta t_0 = 0.2$ s (the smallest $\Delta t$ we evaluate) $N_0 = N_s(\Delta t_0) + N_l(\Delta t_0)$ (Fig 2E). The first obvious feature of Fig 2E is that the fraction of long tracks is >3x larger for circular ($N_{l,R} \approx 0.35$) versus linear ($N_{l,L} \approx 0.12$) DNA across all lag times, indicating that slow modes contribute much more significantly to the transport of rings. The fraction of short tracks for each topology begin to drop off after $\Delta t \approx 0.8$ s, which is the minimum limit we set for acceptable tracks. For circular DNA, $N_l$ surpasses $N_s$ for $\Delta t \gtrsim 1.5$ s, similar to the time at which the downturn in MSD is observed (Fig 1A), indicating that beyond this timescale the slow mode dominates the transport. Conversely, for linear DNA, $N_l$ remains lower than $N_s$ for the full spectrum of lag times, suggesting that slow modes contribute minimally to the linear DNA transport. Moreover, the fraction of total tracks remaining for $\Delta t \gtrsim 1.5$ s is substantially smaller for linear versus circular DNA, with $N/N_0 \approx$ 30-50% and ~55-75% for linear or circular DNA. These data demonstrate that the slow mode dominates the long-time transport of ring DNA, which manifests as nearly halted motion at longer lag times ($\alpha_{l,2} \simeq 0.20$).

To understand the origin of the biphasic behavior of the 'slow' MSD for rings (Fig 2D) we compute the lengthscale at which the crossover in power-law scaling occurs, $r_c \approx \sqrt{\langle (\Delta r(\Delta t_c))^2 \rangle} \approx 450$ nm, which is comparable to the radius of gyration of the circular DNA $R_{G,R} \approx 580$ nm. This finding is a strong indicator that threading of rings by the microtubules underlies the slow mode, as the center-of-mass of a ring pinned by a microtubule cannot move laterally more than $\sim R_{G,R}$. Threading would also restrict the motion along the microtubule contour to below the ~1 µm mesh size of the network, which is ~2-fold larger than $R_{G,R}$. This less stringent constraint along an orthogonal direction is a plausible mechanistic explanation for the relatively faster, yet still strongly suppressed, short-$T$ transport mode that has higher MSD values for all lag times compared to the long-$T$ MSD (Fig 2D).

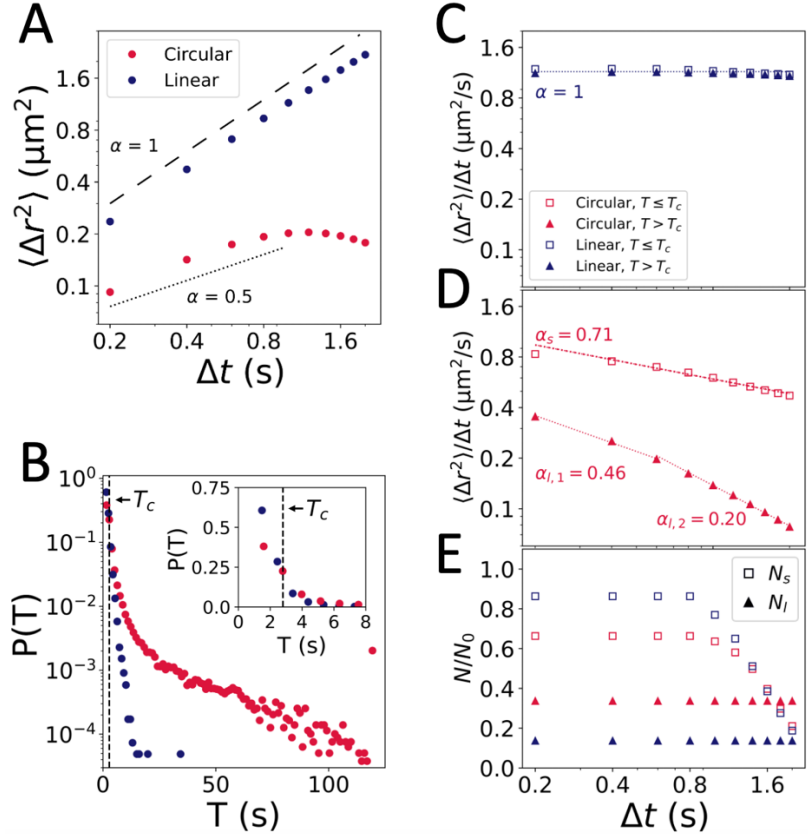

**Figure 2. Microtubule networks induce biphasic subdiffusion of circular DNA but have minimal impact on linear DNA transport.** (A) Mean-squared displacement (MSD), $\langle(\Delta r)^2\rangle = \frac{1}{2}[\langle(\Delta x)^2\rangle + \langle(\Delta y)^2\rangle]$, for linear (blue) and circular (magenta) DNA with dashed and dotted scaling lines corresponding to expected scaling for normal Brownian diffusion ($\alpha = 1$) and representative subdiffusion ($\alpha = 0.5$). (B) Distribution $P(T)$ of durations $T$ that each circular and linear molecule is tracked over the course of each video. The vertical line denotes the critical track length $T_c$ at which $P = P_{max}/e$ for ring DNA. (C, D) MSDs scaled by lag time, $\langle(\Delta r)^2\rangle/\Delta t$, for linear (C, blue) and ring (D, magenta) DNA tracks of length $T < T_c$ (open) and $T < T_c$ (filled). Horizontal regressions correspond to $\alpha = 1$ (normal diffusion) while negative power-law scalings indicate subdiffusion with exponents equating to $\alpha - 1$. Dashed lines are power-law fits to $\langle(\Delta r)^2\rangle/\Delta t \sim \Delta t^{\alpha-1}$ with $\alpha$ values listed. For the $T > T_c$ circular DNA data, which shows clear bimodal scaling, separate fits are performed for $\Delta t = 0.2 - 0.6$ s and $\Delta t = 0.6 - 2$ s, as shown. (E) The number of tracks $N$ with $T < T_c$ ($N_s$, open) and $T > T_c$ ($N_l$, filled), normalized by the total initial number of tracks at $\Delta t_0 = 0.2$ s: $N_0 = N_s(\Delta t_0) + N_l(\Delta t_0)$. As shown, the fraction of long tracks for linear DNA is substantially lower than for rings, and the total number of tracks that remain at the final evaluated lag time is only ~30% for linear DNA compared to ~55% for rings.

**Gaussian transport modes available to linear DNA in quiescent networks are substantially suppressed for rings that display exponential displacement distributions.** To further corroborate and shed further light on the complex transport properties shown in Fig 2, we evaluate the corresponding van Hove distributions $G(\Delta x, \Delta y)$ (Fig 1D) for varying lag times $\Delta t$. As shown in Fig 3A,B, we observe a striking topology dependence, with linear DNA displacements appearing largely Gaussian distributed while the distributions for rings deviate strongly from Gaussianity, displaying sharp peaks at zero and exponential tails at large displacements. Non-Gaussian van Hove distributions are common features of anomalous subdiffusion, as described in the Introduction, and can indeed often be described by a sum of Gaussian and exponential functions [70–73]. Previous studies on transport through confined and crowded systems have shown that the exponential contribution is a result of caging or trapping of molecules (giving the pronounced peaks) as well as hopping between cages (providing the exponential tails), while the Gaussian contribution is from normal Brownian diffusion [5,19,72].

To quantify the different degrees to which Gaussian versus anomalous processes contribute to the dynamics of linear and ring DNA, as evidenced in Fig 3A,B, we fit each van Hove distribution to a sum of a Gaussian and exponential term $G(\Delta d) = Ae^{-\frac{\Delta d^2}{2\sigma^2}} + Be^{-\frac{|\Delta d|}{\lambda}}$, which we find accurately describes the distributions for both topologies across all lag times (Fig 3C-F). However, the relative contributions from the Gaussian and exponential terms to the fits, captured by $A$ and $B$, are strongly dependent on DNA topology across all lag times, which we quantify by evaluating the fractional contribution of the exponential term $b = B/(B + A)$ versus lag time, where $b = 0$ or $b = 1$ describe purely Gaussian or exponential distributions (Fig 3E). As shown, the exponential term comprises >94% of the circular DNA distribution fit across all lag times, with an average value of $\langle b \rangle \simeq 0.97$, while it contributes <40% for linear DNA ($\langle b \rangle \simeq 0.30$). Moreover, for the lag time values in which the majority of the tracks still contribute to the linear data ($\Delta t \lesssim 1.4$ s), the average drops even lower, to $\langle b \rangle \simeq 0.25$. In summary, the presence of passive microtubule networks has a dramatic impact on the transport of ring DNA, leading to extreme non-Gaussianity with pronounced and sharp zero-displacement peaks that are indicative of halted motion, likely arising from threading of rings by microtubules. The absence of these features for linear DNA is further evidence that threading underlies that strongly suppressed transport of rings.

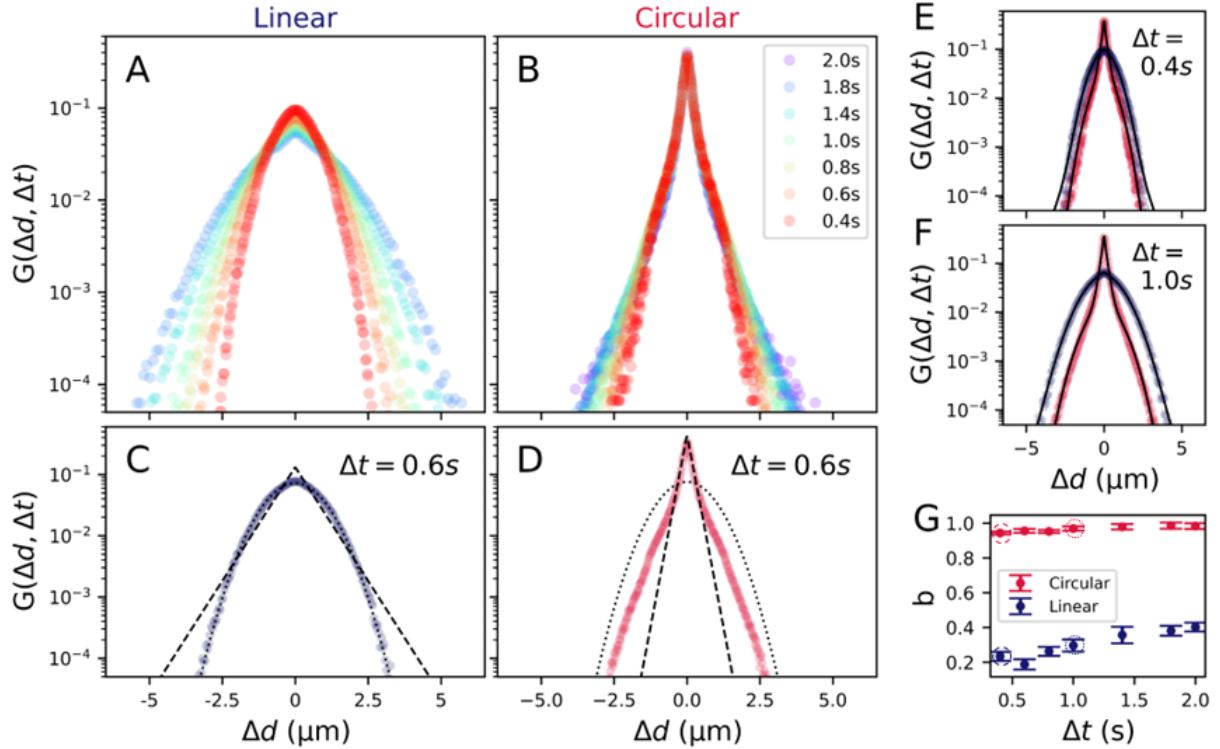

**Figure 3. van Hove distributions for circular DNA are uniquely non-gaussian suggestive of threading by microtubules.** (A, B) van Hove distributions $G(\Delta d, \Delta t)$ where $\Delta d = \Delta x \cup \Delta y$, and $\Delta x$ and $\Delta y$ are the center-of-mass displacements in the $x$ and $y$ directions for linear (A) and circular (B). DNA measured at varying lag times $\Delta t$ is listed in the legend increasing from cool to warm colors. (C, D) $G(\Delta d, \Delta t = \Delta t_c)$ for linear (C) and circular (D) DNA fit to the sum of a Gaussian and exponential $G(\Delta d) = Ae^{-\frac{\Delta d^2}{2\sigma^2}} + Be^{-\frac{|\Delta d|}{\lambda}}$, with the normalized Gaussian term (A = 1, B = 0; dotted) and the normalized exponential term (A = 0, B = 1; dashed) overlaid. (E,F) van Hove distributions and their fits to $G(\Delta d) = Ae^{-\frac{\Delta d^2}{2\sigma^2}} + Be^{-\frac{|\Delta d|}{\lambda}}$ for ring (magenta) and linear (blue) DNA at nominal short and long lag times, $\Delta t = 0.4$ s (E) and $\Delta t = 1$ s (F). (G) Fractional weight of exponential term $b = B/(A + B)$ from fits to van Hove distributions for circular (magenta) and linear (blue) DNA as a function of lag time $\Delta t$. The circled data points correspond to the fits shown in E (dashed) and F (dotted).

**Active network restructuring rescues halted transport of circular DNA while suppressing linear DNA transport.** Armed with an understanding of transport through passive networks, we now investigate the impact of active network restructuring and athermal motion on DNA transport. In particular, we seek to determine if kinesin-driven activity can disrupt the anomalous stalling of DNA rings, and/or more strongly imprint the network onto linear DNA transport; and the extent to which the dynamics of ring and linear DNA can couple to the active network dynamics. To this end, we add ATP to the system to fuel the kinesin clusters to walk along and pull on connected filaments to move and restructure the network. By performing the same analyses as in Fig 2, we first observe a dramatic shift in the MSD for ring DNA, switching from slower than linear chains and extremely subdiffusive ($\alpha \approx 0.2$) without activity (Fig 2A) to faster than linear DNA and ranging from diffusive to superdiffusive ($\alpha \approx$ 1-1.5) with activity (Fig 4A). Conversely, the linear DNA MSD maintains largely diffusive scaling ($\alpha \approx 1$), but with a ~3-fold lower diffusion coefficient compared to the passive case ($D \approx 0.2$ µm² s⁻¹ versus $D \approx 0.6$ µm² s⁻¹). Given the different dynamics we observe for molecules that are tracked for 'long' ($T > T_c$) versus 'short' ($T \leq T_c$) times in the passive case, we evaluate the distribution of track durations for the active networks to assess the extent to which multimodal dynamics may persist (Fig 4B).

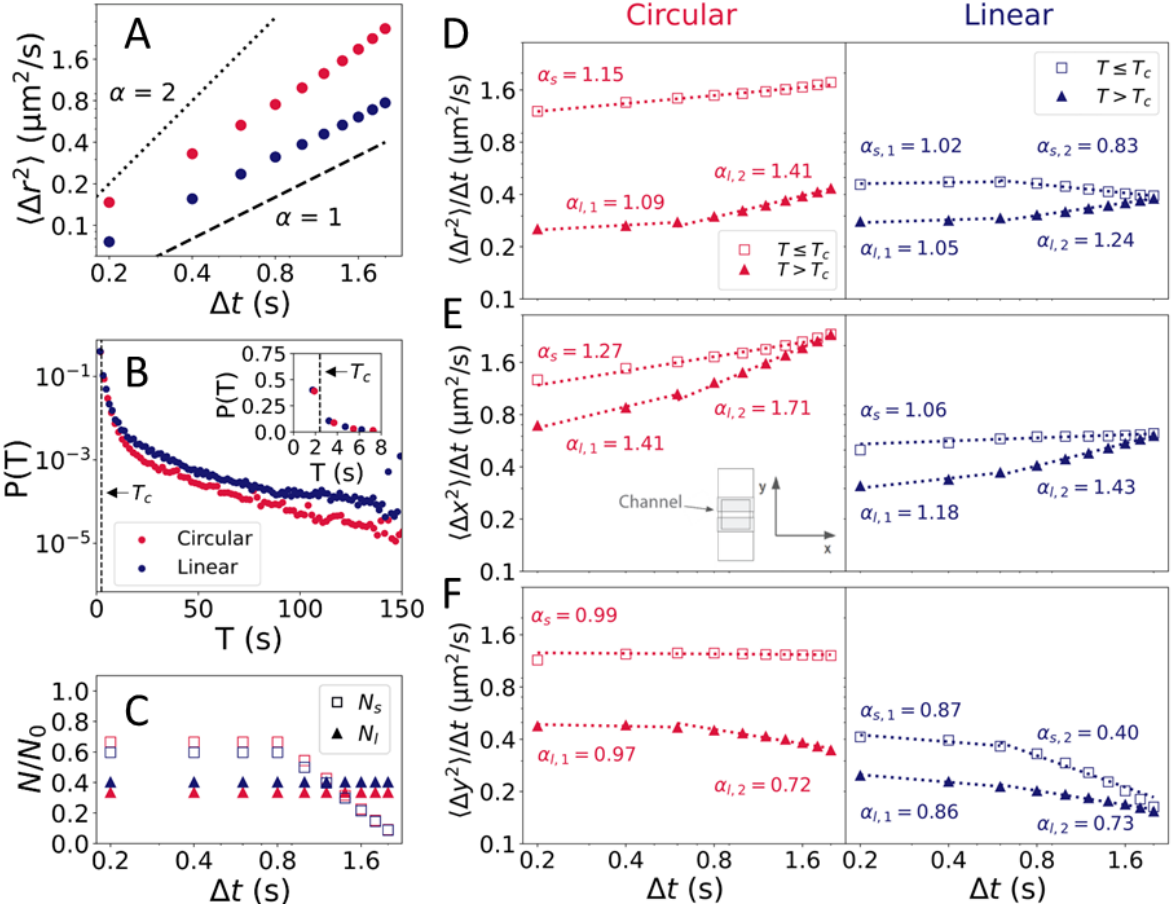

**Figure 4. Network activity facilitates transport of circular DNA while suppressing linear DNA transport.** (A) $\langle(\Delta r)^2\rangle$ versus $\Delta t$ for all tracked circular (magenta) and linear (blue) DNA with dashed and dotted scaling lines corresponding to expected scaling for normal Brownian diffusion ($\alpha = 1$) and ballistic motion ($\alpha = 2$). (B) Distribution $P(T)$ of durations $T$ that each circular (magenta) and linear (blue) molecule is tracked over the course of each video. The vertical line denotes the critical track length $T_c \simeq 2.4$ s at which $P = P_{max}/e$ for ring and linear DNA. Inset shows zoom-in of distributions near $T_c$. (C) The fraction of total tracks $N(\Delta t)/N_0$ with $T \leq T_c$ ($N_s$, open) and $T > T_c$ ($N_l$, filled) as a function of lag time $\Delta t$ shows that the total number of tracks ($N_s + N_l$) remaining at the largest evaluated $\Delta t$ remains >40% for both topologies. (D) $\langle(\Delta r)^2\rangle/\Delta t$ for circular (magenta) and linear (blue) DNA tracks of duration $T \leq T_c$ (open) and $T > T_c$ (filled). (E-F) $\langle(\Delta x)^2\rangle/\Delta t$ (E) and $\langle(\Delta y)^2\rangle/\Delta t$ (F) for circular (magenta) and linear (blue) DNA divided into short-$T$ ($T \leq T_c$, open symbols) and long-$T$ ($T > T_c$, filled) tracks. Dotted lines in D-F are power-law fits of the data to $\sim \Delta t^{\alpha-1}$ with $\alpha$ values listed. For data that show clear bi-phasic scaling, separate fits are performed for $\Delta t \leq \Delta t_c$ (=0.6 s) and $\Delta t \geq \Delta t_c$. For reference, horizontal regressions correspond to $\alpha = 1$ (normal diffusion) while positive and negative power-law scalings indicate superdiffusion and subdiffusion with exponents equating to $\alpha - 1$. Inset in E shows schematic of sample chamber that highlights the large aspect ratio and the orientation of $\hat{x}$ and $\hat{y}$ relative to the chamber dimensions.

As shown in Fig 4B, there are relatively fewer 'long' tracks compared to the passive cases (Fig 2B), and the critical track duration for both linear and ring distributions, where $P = P_{max}/e$, is $T_c \approx 2.4$ s. However, by splitting the data into tracks above and below $T_c$, we observe that the fraction of long tracks $N_l/N_0$ compared to short tracks $N_s/N_0$ is substantially higher for linear DNA compared to the passive case, in line with the slower measured diffusion. $N_l/N_0$ for linear DNA is also modestly higher than that for ring DNA across all lag times, and $N_s/N_0$ is concomitantly lower, in opposition to the passive case. Conversely, activity has minimal impact on the fraction of long and short tracks for rings: long tracks continue to make up a significant fraction of all tracks and dominate the dynamics for $\Delta t > 1.2$ s.

As with the passive case, we examine the MSDs for short and long tracks separately (Fig 4D), to shed light on the intriguing effects of activity on transport that we observe. For ring DNA, similar to the passive case, the MSD for the $T < T_c$ tracks is substantially higher than that for the $T > T_c$ tracks (Fig 4D). Moreover, the $T < T_c$ MSD displays a single power-law scaling while the $T > T_c$ tracks result in bi-phasic scaling of the MSD. The stark difference between the passive and active case for rings is the scaling – shifting from subdiffusive to superdiffusive scaling. Specifically, for $T \leq T_c$ tracks, the scaling exponent shifts from $\alpha_s \simeq 0.7$ (Fig 2D) to $\alpha_s \simeq 1.15$ (Fig 4D), while for $T > T_c$ tracks, the two scaling exponents shift from $\alpha_{l,1} \simeq 0.5$ and $\alpha_{l,2} \simeq 0.2$ (Fig 2D) to $\alpha_{l,1} \simeq 1.1$ and $\alpha_{l,2} \simeq 1.4$. The bi-phasic behavior with similar crossover lengthscale of $r_c \simeq 420$ nm, comparable to $R_{G,R}$, further supports the concept of threading as dictating the long track MSDs for rings in both passive and active cases, which causes more restricted motion at larger lag times in the passive case and more superdiffusive motion in the active. These opposite effects arise from DNA dynamics being strongly coupled to the network dynamics, which are extremely slow in the passive case and ballistic-like in the active case [74,75]. For shorter lag times, the molecules are not fully confined by the network because they are exploring lengthscales $r < r_c$, smaller than both $R_{G,R}$ and $\xi$. The lack of bi-phasic behavior for the short-$T$ tracks, may suggest that the transport mode is only indirectly influenced by the network through, e.g., hydrodynamic interactions, increased viscosity, and/or advection. We explore these possibilities below.

Turning to the linear DNA transport, we observe much closer agreement between MSDs computed for $T \leq T_c$ and $T > T_c$ tracks and scaling that is much closer to diffusive compared to ring MSDs, similar to the passive case. Both MSDs exhibit diffusive scaling for lag times below $\Delta t_c$ then shift to modestly subdiffusive ($\alpha_{s,2} \simeq 0.8$) or superdiffusive ($\alpha_{l,2} \simeq 1.2$) for $T \leq T_c$ and $T > T_c$. The emergence of this weak biphasic behavior indicates stronger coupling to the network. For linear chains, which can entangle with and be confined by the network, but not threaded, the confinement length should be that of the network mesh size, which, if we consider $\langle(\Delta r(\Delta t_c))^2\rangle \approx 0.37$ µm² we compute $r_c \approx 600$ nm, which is ~2-fold smaller than the original nominal mesh size of $\xi \approx 1.1$ µm for the passive network. Given the relation $\langle(\Delta r)^2\rangle \sim D$, this ~2-fold decrease in $r$ should correlate with a 4-fold decrease in $D$ compared to the passive case which is close to the ~3-fold decrease we measure, as described above. For lengthscales larger than $r_c$, DNA transport should couple to dynamics of the network with which it is sterically interacting, which may lead to an increase in $\alpha_{l,2}$, fingerprinting the superdiffusive motion of the active network, while also leading to subdiffusive scaling of $\alpha_{s,2}$ from caging by the smaller mesh size of the contracting network.

**Active and confined transport features are separable by the principal axes of the sample chamber.** To more directly delineate between effects of confinement and active flow, we note that during our experiments, when we observed directed motion of the DNA, it was preferentially in the horizontal $x$ direction of the FOV (see Fig 1C) which corresponds to the long axis of the sample chamber (Fig 4E inset). Specifically, our sample chambers are long narrow cells of dimension 24 mm ($\hat{x}$) × 2 mm ($\hat{y}$) which bias active network motion to be primarily along the long $x$-axis, with minimal directed motion observed along $y$. Thus, to separate effects of active directed motion of the network from altered network structure, we separate the MSDs shown in Fig 4D, i.e., $\langle(\Delta r)^2\rangle = \frac{1}{2}[\langle(\Delta x)^2\rangle + \langle(\Delta y)^2\rangle]$ into their $x$ and $y$ components (Fig 4E,F).

In line with our visual inspection, the MSDs for the two directions are markedly different, with $\langle(\Delta x)^2\rangle$ exhibiting superdiffusive scaling ($\alpha \approx 1.1$-1.7) for both topologies across all lag times and track durations (Fig 4E), while

$\langle(\Delta y)^2\rangle$ exhibits diffusive and subdiffusive scaling ($\alpha \approx 0.4$-$1.0$) (Fig 4F). This anisotropic transport is unique to the active case, whereas $\langle(\Delta x)^2\rangle = \langle(\Delta y)^2\rangle$ in the passive case (Fig S1), demonstrating it is a direct result of motor-driven motion of the network. Moreover, both $\langle(\Delta x)^2\rangle$ and $\langle(\Delta y)^2\rangle$ for rings exhibit weaker confinement compared to linear DNA, with larger MSDs and scaling exponents across all nearly all lag times and track durations, in contrast to passive MSDs that display a nearly opposite trend (Fig 2). These results indicate that motor activity can effectively rescue halted rings by introducing active superdiffusive dynamics into the otherwise rigid network to both facilitate de-threading of rings, tipping the scales from subdiffusive to diffusive, and allow rings that remain threaded to be carried along with the moving network, switching dynamics from extremely subdiffusive to superdiffusive. This tight coupling to the active network dynamics is inaccessible to linear chains that have free ends that prevent threading, resulting in weaker superdiffusivity along $\hat{x}$.

Another feature of $\langle(\Delta x)^2\rangle$ data for both topologies (Fig 4E) that warrants discussion is that long-$T$ tracks display more pronounced superdiffusive scaling, compared to short-$T$ tracks, that becomes steeper beyond $\Delta t_c$. Despite the increased superdiffusivity, which is typically associated with faster motion, the magnitude of $\langle(\Delta x)^2\rangle$ is lower for long-$T$ tracks compared to short-$T$ tracks, indicating more restricted motion. These features together are strong evidence that the transport mode dominating the long-$T$ tracks is steric coupling to the network, which dampens thermal motion and enhances active motion compared to the DNA in solution, leading to smaller MSDs that are more superdiffusive. The higher $\langle(\Delta x)^2\rangle$ values and more pronounced superdiffusion for rings as compared to linear chains ($\alpha_{l,1} \simeq 1.4$, $\alpha_{l,2} \simeq 1.7$ versus $\alpha_{l,1} \simeq 1.2$, $\alpha_{l,2} \simeq 1.4$) further corroborates this physical picture, as threading events are expected to enhance coupling of rings to the network compared to the transient steric entanglements that enable coupling of linear DNA.

The question remains how to understand the mechanism underlying the short-$T$ tracks, which have higher $\langle(\Delta x)^2\rangle$ values, weaker subdiffusion, and the absence of a crossover (i.e., $r_c, \Delta t_c$). Faster, more diffusive and lengthscale-independent dynamics all point to indirect coupling to the activity via advection. Namely, the DNA molecules are primarily interacting with solvent rather than network and are entrained in the advective flow generated by the active network. The smaller size of rings and their reduced ability to entangle with the networks compared to linear DNA facilitate this advective coupling.

Turning to $\langle(\Delta y)^2\rangle$, which we expect to have minimal signatures of active motion based on the sample chamber geometry (Fig 4E) and visual inspection (Fig 1B), we indeed find no superdiffusive scaling (Fig 4F). For long-$T$ tracks, which we suggest are mediated by steric interactions, as described above, we find smaller $\langle(\Delta y)^2\rangle$ values and more subdiffusive scaling compared to the short-$T$ tracks, both indicators of steric confinement by the network and suppressed thermal motion, in line with steric interactions (Fig 4F). We also note that the rings are more weakly subdiffusive than in the passive case (Figs 2D, 4F-left) while linear DNA is more subdiffusive (Figs 2C, 4F-right), which we rationalize as arising from the active restructuring and contraction of the network which allows for more de-threading events (providing more mobility to rings) while decreasing the mesh size (enhancing confinement of the linear chains). Finally, the higher $\langle(\Delta y)^2\rangle$ values and weaker subdiffusion of short-$T$ tracks align with our conjecture that they are dominated by advective or hydrodynamic coupling to the network. This mode manifests as normal diffusion for rings ($\alpha_s \simeq 1$) with a diffusion coefficient $D \simeq 0.5$ µm² s⁻¹ (Fig 4F-left), which is only modestly lower than the free diffusion coefficient measured for linear DNA in the passive case (Fig 2C), indicating only weak influence of the network.

However, unlike the other short-$T$ MSDs, $\langle(\Delta y)^2\rangle$ for linear chains displays a crossover to enhanced subdiffusion at $\Delta t_c$ ($\alpha_{s,1} \simeq 0.9$ versus $\alpha_{s,2} \simeq 0.4$). This effect likely indicates the lengthscale at which the linear chains are able to freely diffuse before feeling the constraints of the network. Only in this case, in which the mesh size is reduced and there is no active motion to counteract the confinement, is this lengthscale small enough to be restrictive over the timescales of our experiments. Specifically, the crossover in $\langle(\Delta y)^2\rangle$ occurs at $r_c \approx 490$ nm (Fig 4I), which is close to our estimated contracted mesh size of ~600 nm.

**DNA is entrained by the heterogeneous ballistic motion of the network due to increased steric confinement.** To corroborate and further characterize the direction-dependent contributions to the transport of both DNA topologies, we evaluate their van Hove distributions of $\Delta x$ and $\Delta y$ separately (Fig 5). We first focus on $G(\Delta y, \Delta t)$, because $\hat{y}$ transport of both topologies appears to lack signatures of active dynamics, so is more readily comparable to the passive case (Fig 3). As shown in Fig 5A,B, we find that both topologies exhibit pronounced non-Gaussian features, similar to the ring DNA data for the passive case (Fig 3B). Moreover, while the linear DNA displacements are largely Gaussian distributed in the passive case (Fig 3A), introducing activity causes $G(\Delta y, \Delta t)$ for the linear DNA to deviate from Gaussianity, even more strongly than the rings (Fig 4A-C). In contrast, activity appears to weaken the extreme non-Gaussianity of rings compared to the passive case (Fig 4D).

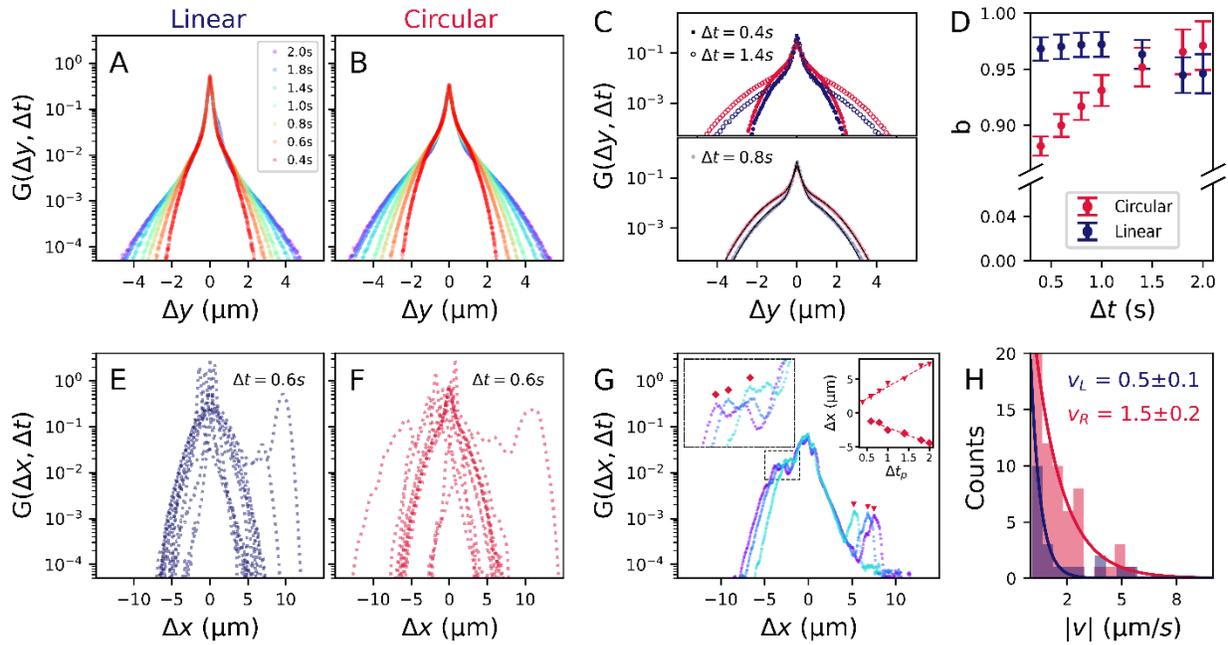

**Figure 5. Kinesin activity induces directed transport and enhanced confinement of DNA dependent on topology and direction.** (A,B) van Hove distributions of center-of-mass displacements in the $y$-direction $G(\Delta y)$ for linear (A) and circular (B) DNA measured at varying lag times $\Delta t$ listed in the legend and decreasing from cool to warm colors. (C) $G(\Delta y)$ distributions and their fits to $G(\Delta y) = Ae^{-\frac{\Delta y^2}{2\sigma^2}} + Be^{-\frac{|\Delta y|}{\lambda}}$ for circular (magenta) and linear (blue) DNA at nominal short and long lag times, $\Delta t = 0.4$ s(open), $\Delta t = 1.4$ s (filled), and $\Delta t = 0.8$ s (bottom). (D) Fractional weight of exponential term $b = B/(A + B)$ from fits for circular (magenta) and linear (blue) DNA as a function of lag time $\Delta t$. (E, F) van Hove distributions of $x$-displacements $G(\Delta x)$ for N = 85, 88 different videos taken across 4 different samples each for linear (E) and circular (F) DNA measured at $\Delta t = 0.6$ s show broad heterogeneity in distributions and non-zero peaks. (G) Sample $G(\Delta x)$ for varying lag times $\Delta t = 0.4$ -1.0 s, colorized as listed in the legend (A), showing the increase in displacement peaks $\Delta x_p$ with increasing $\Delta t$. Left inset shows zoom-in of dotted rectangle region. Right inset shows $\Delta x_p$ versus $\Delta t$ for the marked peaks in G with linear fits to determine corresponding velocities $v = \Delta x_p/\Delta t$. (H) Distribution of measured speeds computed from the distributions shown in E and F for circular (magenta) and linear (blue) DNA with the characteristic speed extracted by exponential fit.

To quantify these topology-dependent effects, we fit all $G(\Delta y, \Delta t)$ distributions to the same sum of Gaussian and exponential terms used in the passive case and evaluate the fractional contribution of the exponential term $b = B/(A + B)$ (Fig 5D). Corroborating our qualitative description, we find that linear DNA distributions are markedly more exponential than for the passive case, with $b > 0.94$ for all lag times, compared to $b < 0.40$ for the passive case (Fig 3G), consistent with the comparatively lower values and subdiffusive scaling of $\langle(\Delta y)^2\rangle$ (Fig 4G). As we describe above, we attribute this increased confinement as arising from the decreased mesh size brought about by network contraction. Specifically, the radius of gyration for the linear DNA, $R_G \approx 960$ nm, is smaller than the nominal mesh size, $\xi \approx 1.1$ µm, of the passive network, but larger than our estimated mesh size of the active network ($\xi \approx 600$ nm), so we expect more pronounced caging and confinement of the linear DNA by the active network. Interestingly, the circular DNA distributions, while still having strong non-Gaussian features, appear to be modestly less confined than in the passive case, with slightly smaller $b$ values at short lag times (0.89 versus 0.95), consistent with the weaker subdiffusion observed ($\alpha \approx 0.97$ versus $\alpha \approx 0.46$). This effect may indicate fewer or shorter-lived threading events, likely due to de-threading facilitated by rearranging filaments.

The $\Delta x$ distributions, which we expect to be sculpted by the active dynamics, display much more complex behavior that cannot be accurately captured by average distributions. Figures 5C and 5D, which shows 12 different $G(\Delta x, \Delta t = 0.6)$ distributions, randomly selected from 85 (Linear) and 88 (Circular) videos collected across four samples each, demonstrates the broad heterogeneity in transport properties. All distributions are distinctly non-Gaussian, and most display signatures of directed motion, as indicated by non-zero peaks. In some cases, the non-zero peaks are smaller secondary peaks while the primary peak (distribution mode) remains at zero, indicating that confinement still dominates the transport but that other active modes also contribute. In other cases, the primary peak is non-zero, demonstrating that directed motion is the primary transport mode (see SI Fig S3-5 for more examples). These active signatures appear to be more pronounced for circular DNA compared to linear, as we expect due to threading events being able to more strongly couple DNA to the network compared to entanglements. Specifically of the 88 different videos collected for each topology, we find 63 instances of active transport features for ring DNA compared to 21 for linear DNA.

If the non-zero peaks indeed arise from ballistic motion, the position of the peaks should scale linearly with lag time and the speed should be the constant of proportionality: $\Delta x = v\Delta t$, as demonstrated in Fig 5G. For a given video, which has a unique $G(\Delta x, \Delta t)$ distribution for each lag time, we find each non-zero peak value $\Delta x_p$ and plot as a function of $\Delta t$ (Fig 5G, Fig 5G left inset). Linear fits to these data yield the speed associated with each set of peaks (Fig 5G right inset). Using this method, we measure a broad range of speeds, with rings displaying more instances of ballistic motion (more counts) with a higher proportionality of large speeds (Fig 5H). Specifically, we find characteristic speeds of $v \simeq 1.5 \pm 0.2$ µm² s$^{-1}$ for ring DNA and $v \simeq 0.5 \pm 0.1$ µm² s$^{-1}$ for linear DNA. Importantly, this range of speeds is within the range of reported cytoplasmic streaming rates of ~1 µm s$^{-1}$, demonstrating physiological relevance; and is also within the reported range of 0.3 – 8 µm s$^{-1}$ for kinesin-driven composites of microtubules and actin [74,75], indicating that the measured transport features indeed map to the active dynamics of the network. Finally, the more instances of ballistic motion and faster corresponding speeds for rings compared to linear DNA are further evidence of threading events that allow rings to more strongly couple to the ballistic motion of the network compared to steric entanglements or advective entrainment.

**Conclusions**

We have shown that macromolecular topology is a subtle yet powerful mechanism to control the coupling of transport phenomena to the surrounding medium. In particular, connecting the two free ends of a macromolecule can dramatically alter the extent to which the dynamics of the surrounding network can imprint onto its transport properties. In cells, there is a constant tension between steric hindrances, crowding and viscoelasticity from the cytoskeleton, organelles, vesicles and soluble proteins, that inhibit macromolecular transport, and enzymatically-driven network restructuring, filament reorganization, and cargo transport that both fight against and leverage these hindrances to allow macromolecules to evade these constraints and sculpt their transport properties, as needed for cellular functions. Yet, understanding the interplay between the seemingly antagonistic contributions to intracellular dynamics remains a grand challenge.

To address this open question we use single-molecule tracking to elucidate the dependence of motor activity and macromolecular topology on the transport of large circular and linear DNA through networks of microtubules and kinesin motors with and without ATP that fuels kinesin activity. Fuelling kinesin introduces two competing effects –athermal network flow and motion that enhances macromolecular transport as well as network contraction that reduces the mesh size and increases confinement effects, thereby suppressing macromolecular transport. Ring DNA molecules leverage the former effect to escape the extreme confinement and halting they experience in passive networks to undergo superdiffusive transport via threading and advection. Conversely, the latter has the most dramatic effect on linear DNA, which is relatively unaffected by the rigid passive network, but is strongly caged by the active network.

Moreover, our results uncover that the remarkably subtle act of closing the ends of a DNA molecule can lead to dramatic and near opposite effects on its transport through cytoskeleton networks, as well as the mechanisms it leverages to couple to the active network dynamics. For example, active network restructuring increases caging and non-gaussian transport modes of linear DNA, while dampening these mechanisms for rings. At the same time, circular DNA exhibits either markedly enhanced subdiffusion or superdiffusion compared to their linear counterparts, in the absence or presence of kinesin activity, providing a direct route towards parsing distinct contributions to transport and determining the impact of coupling on the transport signatures. More generally, leveraging macromolecular topology as a route to programming macromolecular interactions and transport dynamics is an elegant yet largely unappreciated mechanism that cells may exploit to sculpt intracellular transport properties as required for, e.g., streaming, mechanosensing, mixing, transfection, and repair.

**Methods**
**DNA Preparation and Fluorescent Labelling:** Double-stranded 115-kbp DNA is prepared through replication of bacterial artificial chromosomes in *Escherichia coli* followed by purification, extraction, and concentration, as described in Ref. **[76]**. Following purification, supercoiled circular DNA constructs are converted to linear and relaxed circular (ring) topologies via treatment with MluI and topoisomerase (New England Biolabs), respectively. Following enzymatic treatment, DNA solutions are dialyzed into nanopure deionized water and stored at 4°C. Topology, purity and concentration are assessed using gel electrophoresis and UV-vis spectrophotometry.

DNA samples are labeled with MFP488 (MirusBio) following manufacturer protocols to achieve a ~1:5 dye:basepair ratio at a DNA concentration of 100 µg mL$^{-1}$. DNA is added to experimental samples to a final concentration of ~40 ng mL$^{-1}$.

**Kinesin and Microtubule Preparation:** Lyophilized porcine brain tubulin dimers (Cytoskeleton T240) are reconstituted to 5 mg mL$^{-1}$ in PEM-100 and and stored at -80°C in single-use aliquots. Tubulin dimers are polymerized into microtubules and stabilized by adding 8 mM GTP (Cytoskeleton, BST06) and 5 µM paclitaxel (Sigma, T7191) and incubating at 37°C for 30 minutes.

Biotinylated kinesin-401 (bk401) is expressed in T7 Express *lysY* Competent *E. coli* (High Efficiency) (New England Biolabs) and purified as described in Ref. **[74]**. Purified bk401 dimers are suspended in PEM-100 (100 mM PIPES, 2 mM EGTA, 2 mM MgCl2, pH 6.8) supplemented with 0.1 mM ATP and 10% sucrose, then stored at -80°C in single-use aliquots. Dimers of bk401 motors able to bind neighboring microtubules are prepared by incubating bk401 dimers with streptavidin (Thermo-Fisher, FD0663) at a 2:1 molar ratio, supplemented with 166 µM DTT (RPI, 34831223), for 30 minutes at 4°C **[77]**.

**Sample Preparation:** To create experimental sample chambers, a glass coverslip is adhered to a microscope slide via melting of parafilm spacers between them, resulting in ~100 µm thick chambers that accommodate ~10 µL volume. The chambers are filled with 10 mg/mL BSA and incubated for 30 minutes prior to flowing in samples to passivate the surfaces. Polymerized microtubules are mixed with kinesin motor clusters, ATP (RPI, A30030), DNA, and an oxygen scavenging system (glucose, glucose oxidase, catalase) to final concentrations of 6 µM tubulin, 0.22 µM kinesin dimers, 10 mM ATP, 4.8 mM GTP, 6 µM Taxol, 18.8 mg/mL glucose, 2% 2-mercaptoethanol, 17.9 mg/mL glucose oxidase, and 2.9 mg/mL catalase in PEM-100. For the inactive networks, ATP is omitted. Samples are flowed into sample chambers via capillary action and sealed with epoxy.

**Epifluorescence Microscopy Experiments:** Imaging fluorescent-labeled DNA in prepared samples is performed on a home-built inverted epifluorescence microscope. A 488 nm excitation laser (Thorlabs, L488P60) is collimated, spatial-filtered to near TEM00 mode, and expanded. After passing through a tube lens, a longpass dichroic mirror (Thorlabs, MD498) is used to reflect the excitation beam into the objective (Nikon Plan Apo 60x/1.20 WI). The emitted light is collected and passed through the tube lens (Olympus, U-TLU) and a 535 nm bandpass emission filter (Chroma, ET/GFP MCherry) and is imaged onto the camera (Andor, Zyla 5.5 sCMOS). Spatial calibration and verification of field flatness is accomplished using a brightfield source mounted above an objective and a grid slide target (Thorlabs R1L3S3PR).

Videos of labeled DNA embedded in networks are acquired beginning 10 minutes after adding ATP to the sample (for active samples). Videos of at least 600 frames are acquired at 5 frames per second at the mid-plane of the sample in $z$ at different $x, y$ regions of the sample that are separated by at least one field of view. For each DNA topology, >85 videos are acquired over four replicates.

**Single-molecule Tracking:** Using the single-particle tracking Python package TrackPy v0.6.1 **[78]**, we locate the center-of-mass (COM) positions of DNA molecules in the videos and link their COM coordinates $(x, y)$ across frames into trajectories. From these trajectories, we compute COM displacements, $\Delta x$ and $\Delta y$, as a function of lag time $\Delta t$, from which we evaluate time-averaged ensemble mean-squared displacements $\langle (\Delta x)^2 \rangle$, $\langle (\Delta y)^2 \rangle$ and $\langle (\Delta r)^2 \rangle = \frac{1}{2}[\langle (\Delta x)^2 \rangle + \langle (\Delta y)^2 \rangle]$ as functions of lag time $\Delta t$ for $\Delta t = 0.2 - 2$ s. We also construct van Hove probability distributions of particle displacements $G(\Delta x, \Delta y, \Delta t)$ for seven lag times $\Delta t = 0.4$-2 s, determining number of bins using $n_h = 2n^{1/3}$ (Rice's rule). For all analyses, we only consider trajectories that persist for ≥0.8 s (4 frames).

**Acknowledgements:** This work was supported by an NSF RUI Award (DMR-2203791) to J.Y.S., NIH R15 Award (NIGMS R15GM123420) and AFOSR Grant (FA9550-17-1-0249) to RMRA, and NSF DMREF Award (DMR 2119663) to R.M.R.-A. and J.L.R. We are grateful to S.R. and N.S.B. for their analysis of preliminary data sets. We thank A.L., M.A. and N.F. for reagent purification and preparation.

# DNA transport is topologically sculpted by active microtubule dynamics

Dylan P. McCuskey[a], Raisa E. Achiriloaie[a], Claire Benjamin[a], Jemma Kushen[a], Isaac Blacklow[a], Omar Mnfy[a], Jennifer L. Ross[b], Rae M. Robertson-Anderson[c] and Janet Y. Sheung[a]*

[a]Department of Natural Sciences of Scripps and Pitzer Colleges, 925 N Mills Ave, Claremont, CA, 91711, USA.
[b]Syracuse University Department of Physics, Crouse Dr, Syracuse, NY, 13210, USA.
[c]University of San Diego Department of Physics and Biophysics, 5998 Alcala Park, San Diego, CA, 92110, USA.
*jsheung@natsci.claremont.edu

## Supplemental Information

**Figure S1.** DNA transport in inactive networks is isotropic.

**Figure S2.** Determining speeds of linear DNA from non-zero peaks in van Hove distributions.

**Figure S3.** $G(\Delta x, \Delta t)$ computed at $\Delta t = 1.4$ s (blue), 1.8 s (gold), and 2.0 s (green) for each video of circular DNA acquired in one active network sample.

**Figure S4.** $G(\Delta x, \Delta t)$ computed at $\Delta t = 1.4$ s (blue), 1.8 s (gold), and 2.0 s (green) for each video of circular DNA acquired in three other active network samples.

**Figure S5.** $G(\Delta x, \Delta t)$ computed at $\Delta t = 1.4$ s (blue), 1.8 s (gold), and 2.0 s (green) for each video of linear DNA acquired across four active network samples.

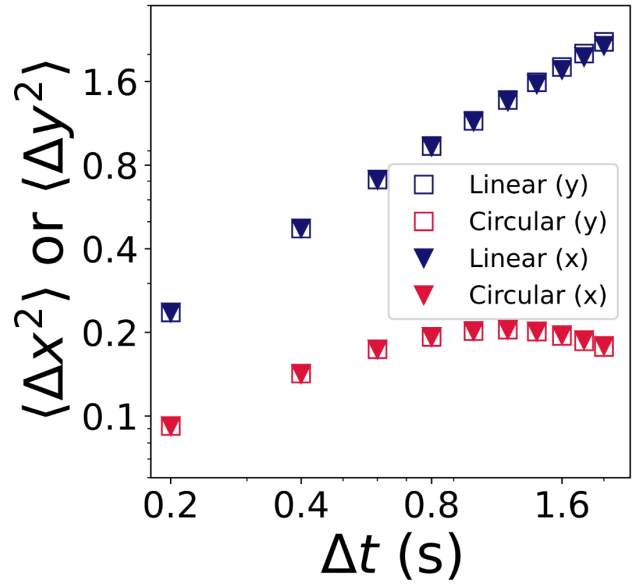

**Figure S1. DNA transport in inactive networks is isotropic.** (A) MSDs for linear (blue) and ring (magenta) DNA, computed separately for $x$ and $y$ directions, $\langle(\Delta y)^2\rangle$ (squares) and $\langle(\Delta x)^2\rangle$ (triangles), are indistinguishable from one another across all lag times $\Delta t$. Data shown is used to compute $\langle(\Delta r)^2\rangle = \frac{1}{2}[\langle(\Delta x)^2\rangle + \langle(\Delta y)^2\rangle]$ plotted in Figure 2A.

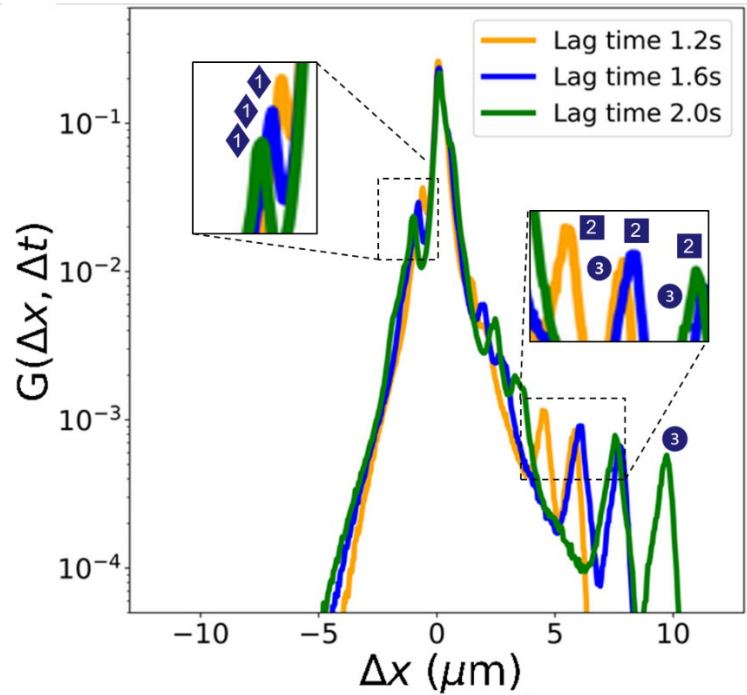

**Figure S2. Determining speeds of linear DNA from non-zero peaks in van Hove distributions.** $G(\Delta x, \Delta t)$ for a sample video of linear DNA, computed for $\Delta t = 1.2$ s (gold), 1.6 s (blue), and 2.0 s (green). Three groups of peaks, numbered 1,2,3 are boxed-in and shown zoomed-in in insets. Each group has a set of three peaks, with peak values $\Delta x_p$ that increase with $\Delta t$. Fitting $\Delta x_p$ versus $\Delta t$ for each group provides a velocity $v$ associated with each group, $\Delta x_p = v\Delta t$. The identical plot is shown for circular DNA in Figure 5G in the main text.

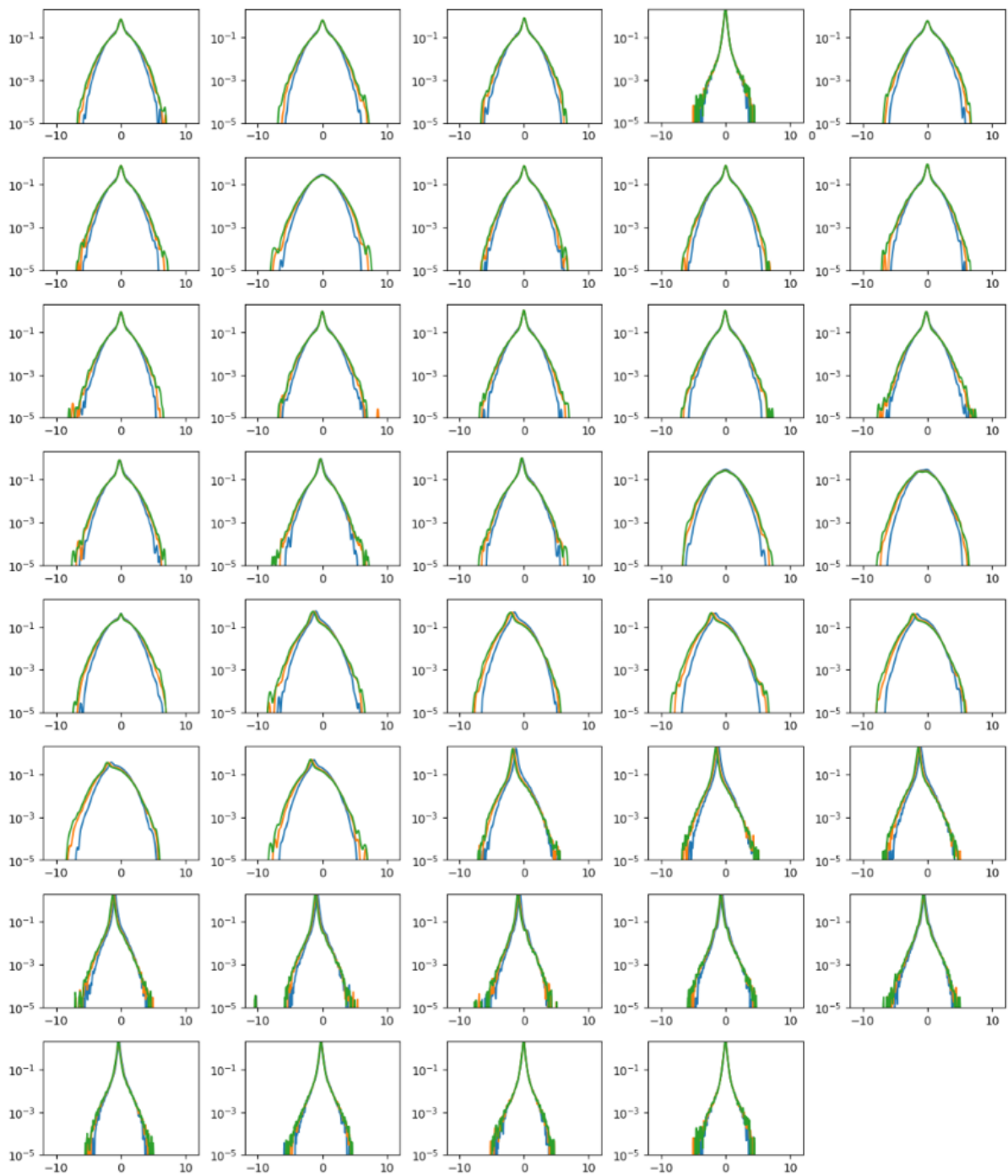

**Figure S3.** $G(\Delta x, \Delta t)$ computed at $\Delta t = 1.4$ s (blue), 1.8 s (gold), and 2.0 s (green) for each video of circular DNA acquired in one active network sample. The y-axis is probability and the x-axis is displacement $\Delta x$ in units of µm.

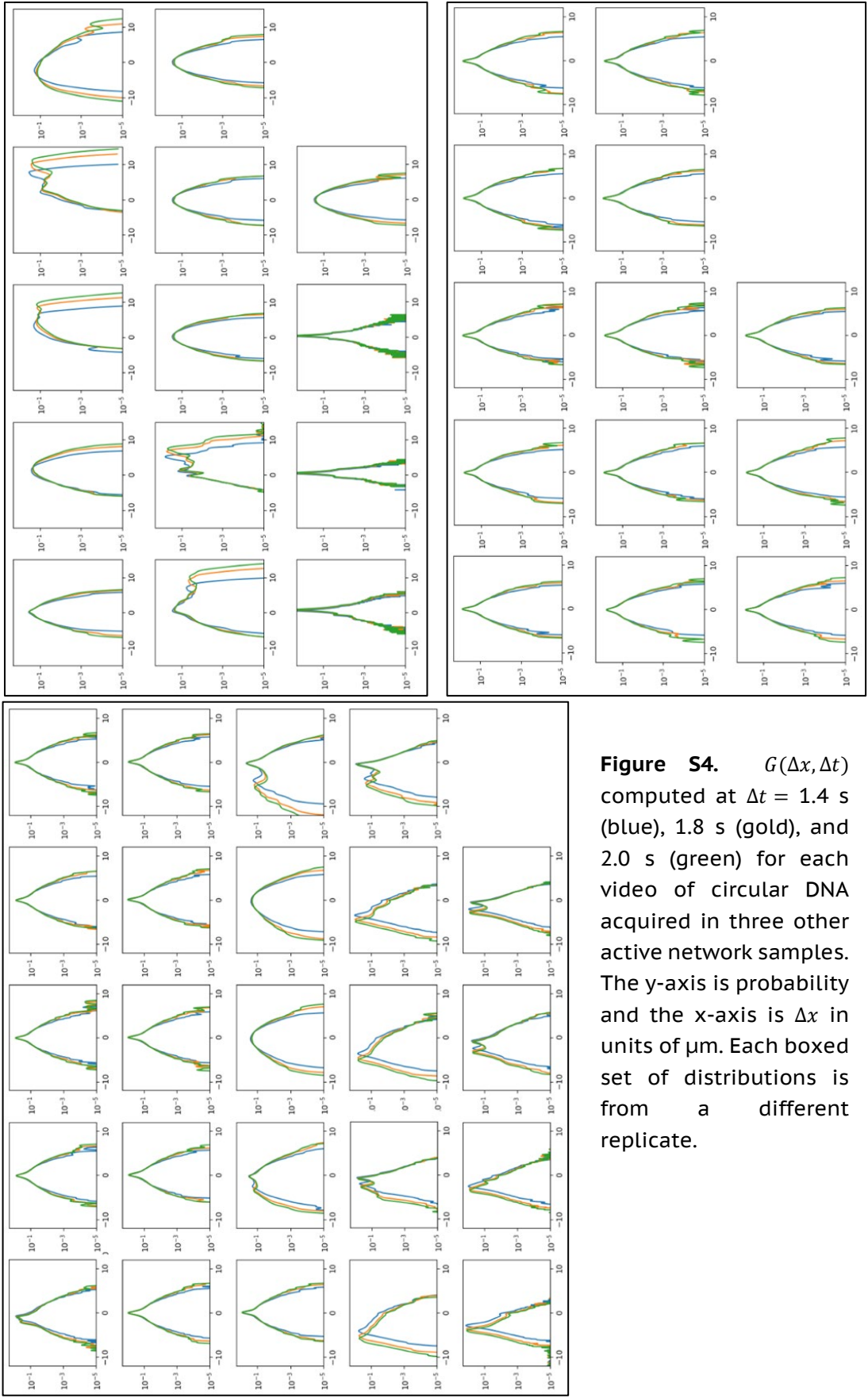

**Figure S4.** $G(\Delta x, \Delta t)$ computed at $\Delta t = 1.4$ s (blue), 1.8 s (gold), and 2.0 s (green) for each video of circular DNA acquired in three other active network samples. The y-axis is probability and the x-axis is $\Delta x$ in units of μm. Each boxed set of distributions is from a different replicate.

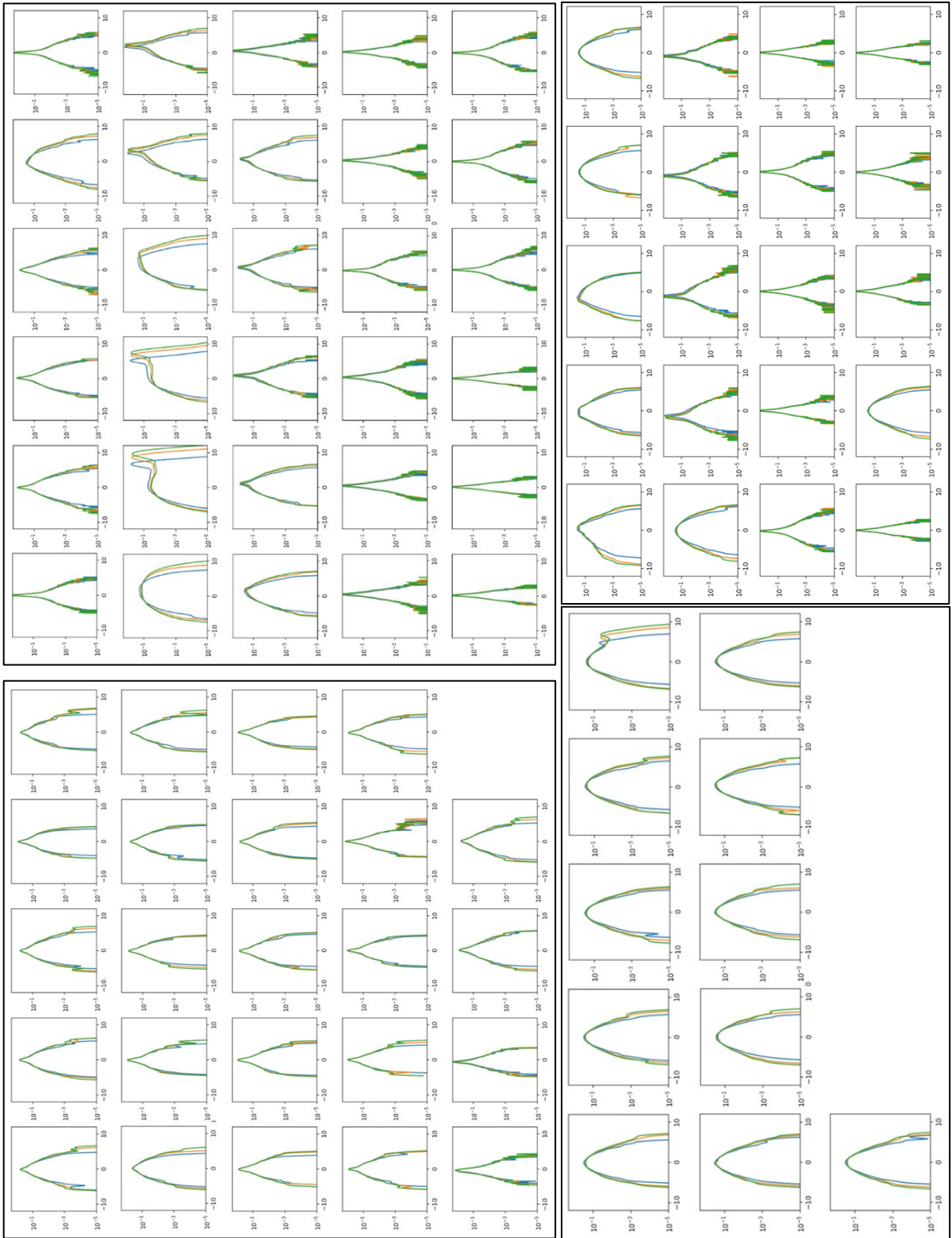

**Figure S5.** $G(\Delta x, \Delta t)$ computed at $\Delta t = 1.4$ s (blue), 1.8 s (gold), and 2.0 s (green) for each video of linear DNA acquired across four active network samples. The y-axis is probability and the x-axis is $\Delta x$ in units of µm. Each boxed set of distributions is from a different replicate.